\documentclass[aps,preprint,superscriptaddress,floatfix]{revtex4-1}

\usepackage{amsmath}
\usepackage{amssymb}
\usepackage{amsfonts}
\usepackage{amsthm}
\usepackage{mathrsfs}
\usepackage{cases}
\usepackage{url}
\usepackage{framed}
\usepackage{cancel,soul}
\usepackage[dvipdfmx]{graphicx}
\usepackage{multirow}
\usepackage[pdftex]{color}
\usepackage{bm}
\usepackage[normalem]{ulem}
\usepackage[colorlinks=true,linktoc=page,linkcolor=blue,citecolor=blue]{hyperref}
\usepackage{gensymb} % for Celcius degrees (\degrees)
\usepackage{siunitx} % for angular degrees (\angle{30})

\def\:={\,\raisebox{0.85pt}{.}\hspace{-2.78pt}\raisebox{2.85pt}{.}\!\!=\,}
\def\=:{\,=\!\!\raisebox{0.85pt}{.}\hspace{-2.78pt}\raisebox{2.85pt}{.}\,}

\begin{document}

\title{
Multiple mobile excitons manifested
as sidebands in quasi-one-dimensional metallic TaSe${}_{3}$}

\author{J.-Z. Ma}
\thanks{Corresponding authors:\\
junzhama@cityu.edu.hk,\,markus.mueller@psi.ch,\,ming.shi@psi.ch}
\affiliation{Department of Physics, City University of Hong Kong,
Kowloon, Hong Kong, China}
\affiliation{Swiss Light Source, Paul Scherrer Institute,
CH-5232 Villigen PSI, Switzerland}
\affiliation{City University of Hong Kong Shenzhen Research Institute, Shenzhen, China}
\author{S.-M. Nie}
\affiliation{Department of Materials Science and Engineering,
Stanford University, Stanford, CA 94305, USA}
\author{X. Gui}
\affiliation{Department of Chemistry, Princeton University, Princeton,
New Jersey 08540, USA}
\author{M. Naamneh}
\affiliation{Swiss Light Source, Paul Scherrer Institute,
CH-5232 Villigen PSI, Switzerland}
\author{J. Jandke}
\affiliation{Swiss Light Source, Paul Scherrer Institute,
CH-5232 Villigen PSI, Switzerland}
\author{C. Y. Xi}
\affiliation{Anhui Province Key Laboratory of Condensed Matter Physics at
Extreme Conditions, High Magnetic Field Laboratory,
Chinese Academy of Sciences, Hefei, Anhui 230031, China}
\author{J. L. Zhang}
\affiliation{Anhui Province Key Laboratory of Condensed Matter Physics at
Extreme Conditions, High Magnetic Field Laboratory,
Chinese Academy of Sciences, Hefei, Anhui 230031, China}
\author{T. Shang}
\affiliation{Key Laboratory of Polar Materials and Devices (MOE), School of Physics and Electronic Science, East China Normal University, Shanghai 200241, China}
\author{Y. M. Xiong}
\affiliation{Anhui Province Key Laboratory of Condensed Matter Physics at
Extreme Conditions, High Magnetic Field Laboratory,
Chinese Academy of Sciences, Hefei, Anhui 230031, China}
\author{I. Kapon}
\affiliation{Department of Quantum Matter Physics, University of Geneva,
24 Quai Ernest-Ansermet, 1211 Geneva, Switzerland}
\author{N. Kumar}
\affiliation{Paul Scherrer Institute, CH-5232 Villigen PSI, Switzerland}
\author{Y. Soh}
\affiliation{Paul Scherrer Institute, CH-5232 Villigen PSI, Switzerland}
\author{D. Gosalbez-Martinez}
\affiliation{Institute of Physics,
\'Ecole Polytechnique F\'ed\'erale de Lausanne (EPFL), CH-1015
Lausanne, Switzerland}
\author{O. Yazyev}
\affiliation{Institute of Physics,
\'Ecole Polytechnique F\'ed\'erale de Lausanne (EPFL), CH-1015
Lausanne, Switzerland}
\author{W. H. Fan}
\affiliation{Beijing National Laboratory for Condensed Matter Physics and
Institute of Physics, Chinese Academy of Sciences, Beijing 100190, China}
\affiliation{University of Chinese Academy of Sciences, Beijing 100049, China}
\author{H. Hubener}
\affiliation{Max Planck Institute for the Structure and Dynamics of Matter,
22761 Hamburg, Germany}
\author{U. De Giovannini}
\affiliation{Max Planck Institute for the Structure and Dynamics of Matter,
22761 Hamburg, Germany}
\author{N. C. Plumb}
\affiliation{Swiss Light Source, Paul Scherrer Institute,
CH-5232 Villigen PSI, Switzerland}
\author{M. Radovic}
\affiliation{Swiss Light Source, Paul Scherrer Institute,
CH-5232 Villigen PSI, Switzerland}
\author{M. A. Sentef}
\affiliation{Max Planck Institute for the Structure and Dynamics of Matter,
22761 Hamburg, Germany}
\author{W.-W. Xie}
\affiliation{Department of Chemistry and Chemical Biology, Rutgers University, Piscataway, New Jersey 08854, USA}
\author{Z. Wang}
\affiliation{Beijing National Laboratory for Condensed Matter Physics and
Institute of Physics, Chinese Academy of Sciences, Beijing 100190, China}
\affiliation{University of Chinese Academy of Sciences, Beijing 100049, China}
\author{Christopher Mudry}
\affiliation{Condensed Matter Theory Group, Paul Scherrer Institute, 
CH-5232 Villigen PSI, Switzerland}
\affiliation{Institute of Physics,
\'Ecole Polytechnique F\'ed\'erale de Lausanne (EPFL), CH-1015
Lausanne, Switzerland}
\author{M. M\"uller}
\thanks{Corresponding authors:\\
junzhama@cityu.edu.hk,\,markus.mueller@psi.ch,\,ming.shi@psi.ch}
\affiliation{Condensed Matter Theory Group, Paul Scherrer Institute, 
CH-5232 Villigen PSI, Switzerland}
\author{M. Shi}
\thanks{Corresponding authors:\\
junzhama@cityu.edu.hk,\,markus.mueller@psi.ch,\,ming.shi@psi.ch}
\affiliation{Swiss Light Source, Paul Scherrer Institute,
CH-5232 Villigen PSI, Switzerland}
%\date{\today}

\begin{abstract}
\textbf{
Charge neutrality and an expected itinerant nature makes excitons
  potential transmitters of information. However, exciton mobility
  remains inaccessible to traditional optical experiments that only
  create and detect excitons with negligible momentum. Here, using
  angle-resolved photoemission spectroscopy, we detect dispersing
  excitons in quasi-one-dimensional metallic trichalcogenide,
  TaSe$_3$. The low density of conduction electrons and low
  dimensionality in TaSe$_3$ combined with a polaronic renormalization
  of the conduction band and the poorly screened interaction between
  these polarons and photo-induced valence holes leads to various
  excitonic bound states that we interpret as intrachain and
  interchain excitons, and possibly trions. The thresholds for the
  formation of a photohole together with an exciton appear as side
  valence bands with dispersions nearly parallel to the main valence
  band, but shifted to {\em lower} excitation energies. Interestingly,
  the energy separation between side and main valence band can be
  controlled by surface doping, enabling the tuning of certain exciton
  properties. } 
\end{abstract}

%75.10.Pq 	Spin chain models
%64.70.Tg 	Quantum phase transitions
\maketitle
%\tableofcontents

\textbf{Introduction}
Low-dimensional electronic systems with a low density of charge carriers
$n^{\,}_{\mathrm{c}}$ exhibit interesting many-body effects.
Indeed, their Fermi energy is low as compared to the strength
of the Coulomb interaction, which is poorly screened. 
Moreover, the low dimensionality enhances interaction effects.
They manifest themselves in 
non-Fermi liquid gapless ground states
such as metallic Luttinger liquids
or gapped ground states such as insulating Wigner
crystals and other kinds of charge density waves
\cite{Nagaosa99quantum,Giamarchi04}.
Not only the ground state of such strongly interacting systems,
but also their excitations exhibit interesting strong coupling phenomena.
In particular, the attraction between negatively charged electrons
and positively charged holes can lead to bound states
in the excitation spectrum, usually referred to as excitons%
~\cite{Frenkel31,Wannier37,Kasha59,Knox83,Combescot16,Wang18},
and this tendency is enhanced in low dimensions.

The creation in insulators
of \textit{non-moving} excitons [that is,
bound states from electrons and holes
located at the minimum and maximum of the conduction band (CB)
and valence band (VB), respectively] by optical excitation
is fairly standard \cite{Kasha59,Knox83,Wang18}.
Very recently, distinct branches of \textit{dispersing} excitons have
been observed in free-standing monolayer WSe${}_{2}$, a
two-dimensional Dirac band insulator, by momentum-resolved electron
energy-loss spectroscopy \cite{Hong20}. However,
the observation of mobile bound states
with sharp dispersions in \textit{metals} has remained elusive for
various reasons.
Higher-dimensional metals are unlikely to host excitons.
First, their creation requires an exceedingly strong {(despite screening)}
Coulomb interaction that is not preempted by an interaction-driven
instability to an insulator. Second,
the excitation of moving excitons with light involves
higher-order processes, whose cross-sections are very weak, unless
low dimensionality and low carrier density (as present in
TaSe${}_{3}$) allow for strong Coulomb interactions with small
momentum transfer. The heavy quasiparticles (polarons)
constitute a third favorable attribute of TaSe${}_{3}$,
as their mass increases the binding energy of the resulting excitons
(see Supplemental Material for a more
detailed discussion).

The interplay of dilute conduction electrons, low effective
dimensionality, and heavy quasiparticles in TaSe${}_{3}$ makes this
Q1D material a prime candidate for probes of excitonic effects by
ARPES.

%\sout{\textbf{Main experimental results}}
We have systematically studied the metallic phase of TaSe${}_{3}$
using ARPES and observed the following features at low temperature.  
(1) Several side-valence bands (SVBs) appear
\textit{exclusively above} a pronounced VB
-- in contrast to most ARPES spectra that report sidebands
(Figs.\ S1-S3 of the Supplementary Material).
(2) Their dispersions are roughly
parallel to the VB.
(3) When  the surface carrier density is increased by doping,
the energy separations between the SVBs increase.
(4) Close to the Fermi level, the CB is heavily renormalized,
and the coherent quasiparticle peak follows a $\mathsf{W}$-shaped dispersion.
As we will argue, observations (1-4)
suggest that the SVBs result from
strong coupling between the valence and conduction electrons and
involve \textit{mobile} bound states (excitons, and perhaps trions)
that have not been observed so far using ARPES.
Up to now, ARPES has detected excitonic physics only 
though the effects of
excitonic condensation, as seen for example, in the
electronic structure of Ta${}_{2}$NiSe${}_{5}$ near the Fermi level
\cite{Seki14}, or through the band
folding due to a finite momentum condensate, for example,
in 1T TiSe${}_{2}$ \cite{Sugawara15}.

\textbf{Results}

\begin{figure*}[t]
\begin{center}
\includegraphics[angle=0,width=0.9\textwidth]{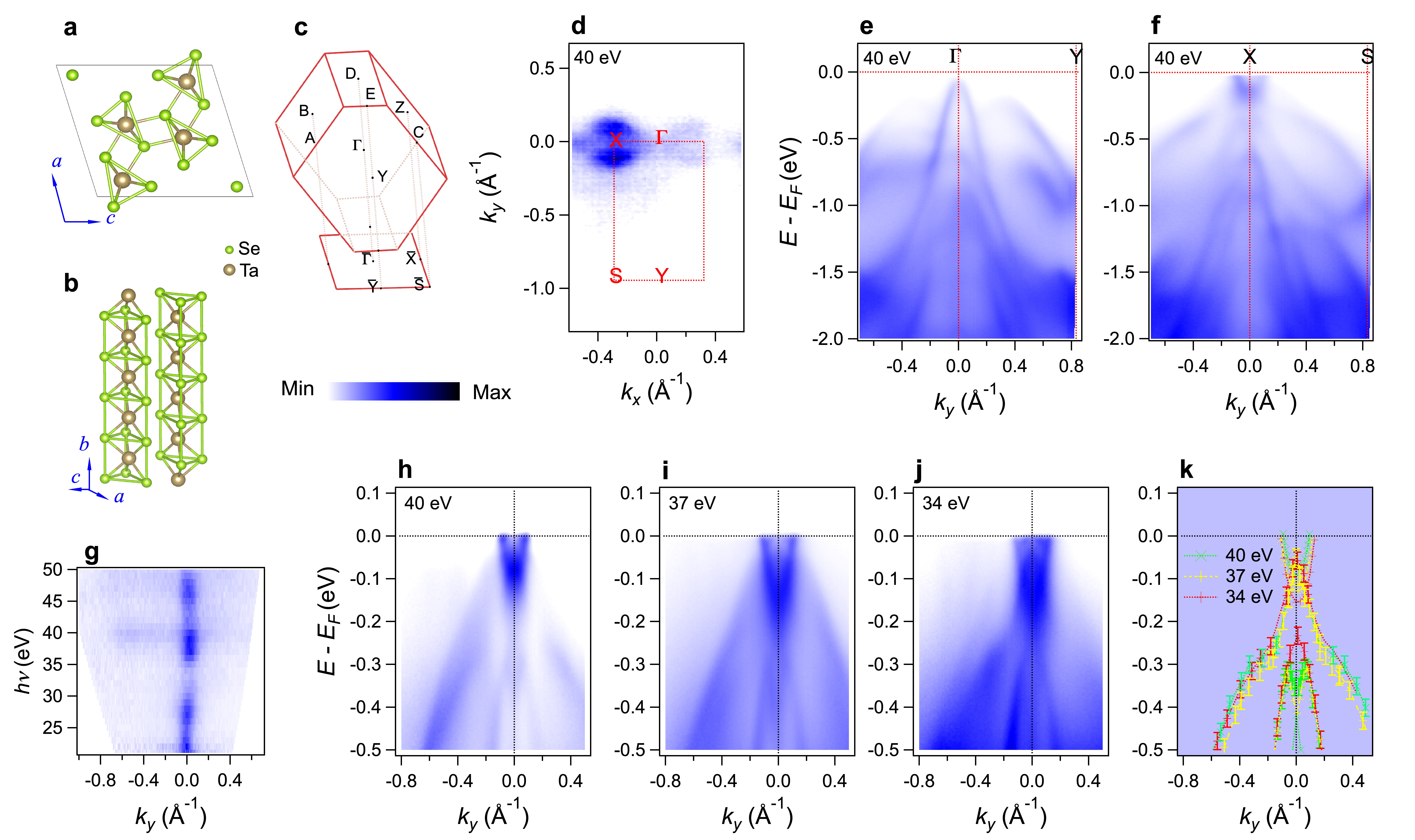}
\end{center}
\caption{%
The crystalline and electronic structures of TaSe${}_{3}$ at
$\mathrm{15K}$. 
\textbf{a} Crystal structure projected onto the $a$-$c$ plane. The crystallographic axis are $a$, $b$, and $c$.
\textbf{b} Crystal structure with
its 1D chains oriented along the $b$ axis.
\textbf{c} The bulk Brillouin zone BBZ
and the 2D surface Brillouin zone SBZ
of the cleavage surface [normal to $(10\bar{1})$].
\textbf{d} ARPES intensity at the Fermi energy $E^{\,}_{\mathrm{F}}$
as a function of the momentum
$\bm{k}\in\mathrm{SBZ}$
for incoming photon energy $h\,\nu=40\,\mathrm{eV}$.
The momenta
$k^{\,}_{x}$ and $k^{\,}_{y}$
are conjugate to the (101) and the $(010)$ direction, respectively.
\textbf{e-f} ARPES intensities at $h\,\nu=40\,\mathrm{eV}$ as a function of
the energy $E-E^{\,}_{\mathrm{F}}$ and the momentum component $k^{\,}_{y}$
in the SBZ, along the cut $\Gamma$-Y ($k^{\,}_{x}=0$) in \textbf{e},
and along the cut X-S in \textbf{f}.
\textbf{g}
ARPES intensity as a function of the momentum $k^{\,}_{y}$ and of the
incoming photon energy $h\,\nu$, which tunes the momentum $k^{\,}_{\perp}$
perpendicular to the planes. The intensity contour is
seen to be very elongated along $k^{\,}_{\perp}$, suggesting that the equal energy
contour near Fermi level is essentially an elliptical cylinder with axis parallel to
$k^{\,}_{\perp}$.
\textbf{h-j} ARPES intensities as a function
of the energy $E-E^{\,}_{\mathrm{F}}$ and the momentum $k^{\,}_{y}$
along the cut X-S in the SBZ
for photon energies 
$h\,\nu=40\,\mathrm{eV}$ \textbf{h},
$h\,\nu=37\,\mathrm{eV}$ (i),
and
$h\,\nu=34\,\mathrm{eV}$ \textbf{j}.
\textbf{k} Superimposed band dispersions from panels \textbf{h-j}
(shown there as colored continuous lines as guide to the eye).
The variation with $h\,\nu$ implies a finite dispersion along $k^{\,}_{\perp}$.
        }
\label{Fig: 1}
\end{figure*}

\textbf{Material characterization}
The trichalcogenide TaSe${}_{3}$ consists of covalently bonded layers,
stacked and held together by weaker van der Waals forces along the direction
$(10\bar{1})$ \cite{Srivastava92,Island17}. Each layer
consists of chains oriented along the $b$-axis, see Figs.\
\ref{Fig: 1}\textbf{a-b},
with strongly anisotropic electric and optical response.
The natural cleavage plane is the $(10\bar{1})$
surface. The corresponding bulk Brillouin zone (BBZ) and the
$(10\bar{1})$ surface Brillouin zone (SBZ) are shown in
Fig.\ \ref{Fig: 1}\textbf{c}.
In the SBZ,
$(k^{\,}_{x}, k^{\,}_{y})$
denote the components of momentum along $(101)$ and $(010)$,
respectively.

TaSe${}_{3}$ is so far the only known trichalcogenide
that is a metal at high temperature and becomes
superconducting below $\mathrm{2K}$ without forming a charge density wave (CDW)
\cite{Sambongi77,Tsutsumi77,Yamamoto78,Haen78,Ekino87,Cava85,Nagata89}.
However, it is close to such a transition,
as suggested by CDW-like signatures observed upon
doping with Cu \cite{Nomura2017,Nomura2019},
application of strain, Ref.\ \cite{Yang2019}
and our own surface doping results reported below.
We note that topological surface states\, \cite{Nie18} have been observed 
in Refs.\ \onlinecite{Chen2020,Lin2020}
in the gap between the top of the VB and the bottom
of the CB of TaSe${}_{3}$,  where band inversion occurs.
However, as we will argue, they are clearly distinct from the excitonic
features we focus on here.
 
\textbf{Overall band structure from ARPES}
We have studied the electronic
structure of TaSe${}_{3}$ with ARPES on the \textit{in situ} cleaved surface
$(10\bar{1})$.
The intersection of the small three-dimensional
(3D) Fermi surface (FS)
with a plane spanned by momenta conjugate to
the crystalline directions $b$ and $a+c$
is seen as the region of highest intensity
in Fig.\ \ref{Fig: 1}\textbf{d}.
Figure\ \ref{Fig: 1}\textbf{e} ($\Gamma$-Y cut) shows hole-like VBs
with maxima at the zone center $\Gamma$, dispersing from
$0.1\,\mathrm{eV}$ to $2\,\mathrm{eV}$.
Figure \ref{Fig: 1}\textbf{f} (X-S cut)
shows an electron-like CB with minimum at the
mid-point X on one zone edge of the BBZ. The strong anisotropy in the
ac plane results in a FS in the form of an elongated elliptical
electron pocket centered at X, see Fig.\ \ref{Fig: 1}\textbf{d}, 
the minor axis along
the X-S direction reflecting the strong dispersion along the $b$-axis.
There is no good nesting wave vector for the FS of TaSe${}_{3}$ shown
in Fig.\ \ref{Fig: 1}\textbf{d}. This might rationalize
the absence of a CDW phase in pristine TaSe${}_{3}$.
The coupling between stacked layers is weak.
This is seen both by the elongated FS in a 
cut with the plane $k^{\,}_{x}=0$, Fig.\ \ref{Fig: 1}\textbf{g}
(the photon energy $h\,\nu$ being used to explore momenta normal
to the cleavage plane),
as well as in the similarity of the spectra in Figs.\ \ref{Fig: 1}\textbf{h-j},
corresponding to different normal momenta.
Nevertheless, Fig.\ \ref{Fig: 1}\textbf{k}
shows a definite, if small, dependence
on the incoming photon energy, which we take as evidence for these
bands being bulk as opposed to surface bands.

\begin{figure*}[t]
\begin{center}
\includegraphics[angle=0,width=0.9\textwidth]{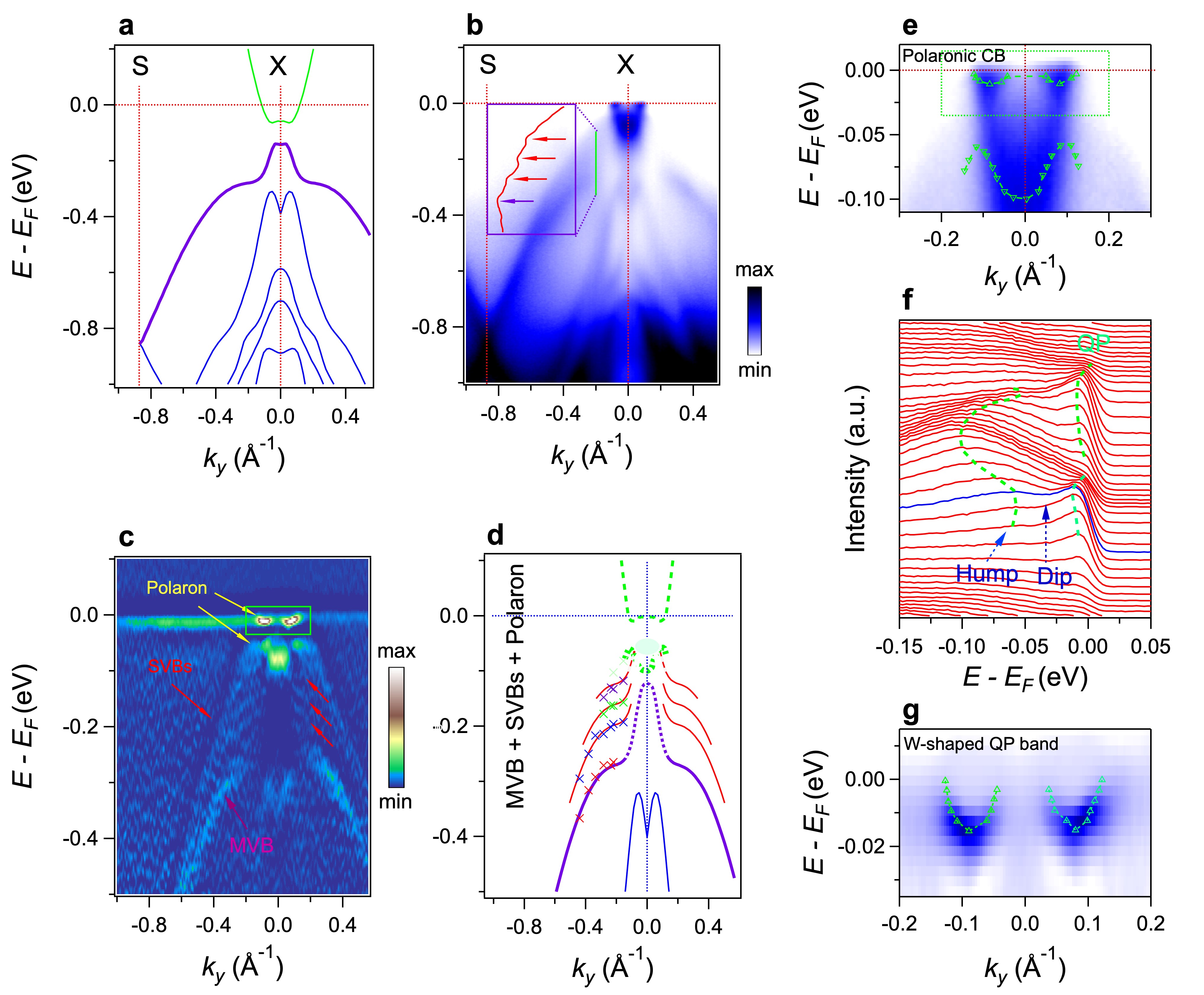}
\end{center}
\caption{%\footnotesize
Side valence bands and polarons in pristine sample.
\textbf{a} Dispersion of the noninteracting (Bloch) bands of TaSe${}_{3}$
computed using DFT (see  method part).
The purple band is the topmost VB, which we refer to as
the MVB. It lies between one CB ({colored in} green)
and four lower lying valence bands ({colored in} blue).
The dispersion of all bands is symmetric about X along the X-S cut.
\textbf{b} ARPES data from Fig.\ \ref{Fig: 1}\textbf{f}
with higher energy resolution and statistics.
The curve in the inset shows the EDC along the green line
($k^{\,}_{y}=-0.2\,\text{\AA}^{-1}$).
The red arrows point to side peaks.
Their dispersion{s} with momentum define
``side valence bands'' (SVBs).
\textbf{c} The curvature intensity plot corresponding to the data in \textbf{b}
enhances the visibility of SVBs marked by red arrows above the MVB.
The green box 
encloses the coherent branch of the polaron quasiparticles (QPs)
making up the bottom of the CB and whose spectral weight
is concentrated in two small islands of diameter $\Delta k$.
The broad hump indicated by a yellow arrow is
the remnant of the CB that would be left,
were there no interaction with a bosonic mode.
\textbf{d} {Schematic} interpretation of the signatures seen in \textbf{c}.
\textbf{e-f} Close up of the data of Fig.\ \ref{Fig: 1}\textbf{b}.
The upper and lower dashed lines in \textbf{e}
trace the dispersions of the polaron QP
peak and the broad hump, respectively, as visible in the EDCs in \textbf{f}.
\textbf{g} Detailed view of the ARPES intensity within the polaron band
of in the green box of \textbf{c}.
The dashed green line indicates the extrapolated dispersion
of the $\mathsf{W}$-shaped QP band.
       }
\label{Fig: 2}
\end{figure*}

\textbf{Side-valence bands (SVBs) }
Here, we focus our attention on SVBs, which are remarkable
spectral features visible in the energy distribution curves (EDC) of
the ARPES spectra.
We show in
Fig.\ \ref{Fig: 2}\textbf{a} the band structure of
TaSe${}_{3}$ along the X-S direction in the BBZ
predicted by density functional theory (DFT),
as explained in the Methods section.
There is one CB (green),
lying above five VBs. The VB closest to the CB
(purple) is referred
to as the main-valence band (MVB).
The lower lying VBs are colored in blue.
These DFT bands match the five bands measured in
Fig.\ \ref{Fig: 2}\textbf{b}.
The EDC along the vertical green line with fixed
$k^{\,}_{y}\approx0.2\,\text{\AA}^{-1}$
in Fig.\ \ref{Fig: 2}\textbf{b} is shown in the boxed inset.
The arrows indicate local maxima of this EDC.
They move as $k^{\,}_{y}$ varies, defining SVB dispersions.
The SVBs are bounded from below by the MVB
which is stronger in intensity.
The MVB and the SVBs become better visible in the curvature intensity plot
of Fig.\ \ref{Fig: 2}\textbf{c} associated with the data of
Fig.\ \ref{Fig: 2}\textbf{b},
(the color scale being related to the curvature of the ARPES intensity).
For reference, Fig.\ \ref{Fig: 2}\textbf{d} shows
a schematic of the band dispersion in Fig.\ \ref{Fig: 2}\textbf{c}.
The dispersions of the MVB (purple arrow) and three SVBs (tilted red arrows)
are approximately symmetric about X.
The Supplementary Material shows more EDCs (Figs.\ S4-S7)
illustrating how the SVBs are identified and evolve with doping.
We only observe sidebands of the MVB,
but none associated with the CB, or the deeper VBs.
Close to the X point, there is no observable intensity that could be
clearly assigned to either SVBs or to the MVB.
In Fig.\ \ref{Fig: 2}\textbf{d}
the evolution of the MVB away from large momenta (where its
peak is well resolved) is indicated by a dashed segment, delineating
the ``nose'' predicted by DFT. We also indicate the possible
continuation of SVBs as they approach the spectral features
associated to the CB, colored in green in both
Fig.\ \ref{Fig: 2}\textbf{c} and Fig.\ \ref{Fig: 2}\textbf{d}.
The latter are of a polaronic origin and play an important role
for the excitonic features seen in ARPES.

\textbf{Polaron band}
In typical members of the family XT${}_{3}$,
such as NbSe${}_{3}$ \cite{Ekino87},
FS nesting induces a CDW instability at fairly high temperatures,
pointing toward a substantial electron-phonon coupling
in these materials.

In pristine TaSe${}_{3}$, the conduction electrons give rise
to two branches of excitations, as traced by the maxima of the
EDCs in Figs.\ \ref{Fig: 2}\textbf{e} and \ref{Fig: 2}\textbf{f}.
The   branch of ``bare'' electron excitations
yields a parabolic CB with a fairly short life time,
showing up as a broad hump as a function of energy. Much sharper,
phonon-dressed quasiparticle (QP) excitations form a strongly
renormalized, weakly dispersive polaron branch close to the Fermi
energy, separated by a dip from the hump of bare excitations.  The
spectral weight of the heavy polaron band is strongest close to
$k^{\,}_{\mathrm{F}}$ and becomes very weak for $k^{\,}_{y}\approx
0$. Furthermore the dispersion of the maximal spectral weight is not
essentially flat as expected for a simple heavy polaron, but
follows a non-monotonic $\mathsf{W}$ shape, see Fig.\ 2\textbf{g}.
Such a dispersion might arise due to the  hybridization of the polaron band
with a dispersing composite excitation consisting,
e.g., of a conduction electron
and a nearly soft phonon. 
We do not explore these features in more detail here, but
simply observe that they entail that the heavy polaron band effectively breaks
up into two small islands concentrated just below the Fermi points
$\pm k^{\,}_{\mathrm{F}}$. This is important in the context of 
exciton formation and SVBs in ARPES. 

\textbf{Effects of surface doping}
Crucial insight into the nature of the SVBs
is gained by increasing the density
of surface electrons. This is achieved by evaporating potassium (K)
in situ on the cleaved surface of TaSe${}_{3}$.
Here we  focus on moderate doping
(evaporation times of one minute or less),
while stronger doping data (up to five minutes of evaporation)
are reported in Fig.\ S9
of the Supplementary Material.

Figures \ref{Fig: 3}\textbf{a} and \ref{Fig: 3}\textbf{b} 
show the ARPES intensity after evaporation for
$\hbox{t1}\equiv\hbox{1\,minute}$,
for the same range of momenta and energies as in Fig.\ \ref{Fig: 2}\textbf{b}.
The evaporated atoms chemically dope electrons onto the surface
which increases the filling of the conduction band.
The incoming photons in Fig.\ \ref{Fig: 3}\textbf{a}
are circularly polarized with positive helicity,
while they linearly polarized in the incident plane in Fig.\
\ref{Fig: 3}\textbf{b}
(the scattering geometry is reported in Fig.\ S8).
EDCs, such as shown in Fig.\ \ref{Fig: 3}\textbf{c} (or Fig.\ S7),
allow to locate the MVB and the SVBs.
More EDCs for the same evaporation time t1 can be found in 
Figs.\ S5 and S6 of the Supplementary Material.
The curvature intensity plot in Fig.\ \ref{Fig: 3}\textbf{d} shows 
that the two SVBs (indicated by the red arrows)
are more pronounced than in the undoped case of Fig.\ \ref{Fig: 2}\textbf{c}.
A heavy, but non-monotonically dispersing polaron band
close to the Fermi energy is still present, but the size of the high intensity
islands of diameter $\Delta k$  
within the polaron band increases together with $k^{\,}_{\mathrm{F}}$,
as is apparent in Figs.\ \ref{Fig: 3}\textbf{d}.
The measured dispersions are traced in
Fig.\ \ref{Fig: 3}\textbf{e}.
Note that the non-renormalized CB branch 
is essentially parabolic, apart from the polaronic effects close
to the Fermi energy and the avoided level crossings with some SVBs,
where the ARPES intensity is suppressed.
For large enough momenta one identifies two SVBs roughly parallel
to the MVB. At small momenta close to X,
there are two nose-like dispersing pieces of an excitation branch.
They might be continuations of the SVB's seen at larger momenta,
as hinted in Fig.\ \ref{Fig: 3}\textbf{f}.

Figure\ \ref{Fig: 3}\textbf{g} shows the ARPES intensity at
$h\,\nu=37\,\mathrm{eV}$ after doping. Here, the first SVB is clearly
visible.

The doping strongly affects the average energy separations
$\Delta^{\,}_{1}$ and $\Delta^{\,}_{2}$
between the MVB and the first SVB, and the first and second SVBs
(if detectable),
respectively. We measure these spacings in the regime of large momenta
relative to X. In Fig.\ \ref{Fig: 3}\textbf{h},
$\Delta^{\,}_{1}$ and $\Delta^{\,}_{2}$ are seen to substantially increase
with doping or, equivalently, with the Fermi wavevector $k^{\,}_{\mathrm{F}}$.
In particular,
$\Delta^{\,}_{1}\sim180\,\mathrm{meV}$ and 
$\Delta^{\,}_{2}\sim70\,\mathrm{meV}$,
in the presence of potassium doping (t1) are 
much larger than the spacings $\Delta^{\,}_{1}\sim70\,\mathrm{meV}$ and
$\Delta^{\,}_{2}\sim37\,\mathrm{meV}$ in undoped TaSe${}_{3}$.

Effects of stronger doping 
are presented in Figs.\ S9-S10 of the Supplementary Material.
The QP-dip-hump features at the Fermi energy described above
evolve into a CDW at $\hbox{t2}\equiv\hbox{2\,minutes}$.
This observation supports the interpretation of these polarons
as originating from an electron-phonon coupling in TaSe$_3$.
More emperical evidences for electron-phonon coupling in TaSe$_3$
is provided by the evolution of Raman
intensities in Fig.\ S12 of the Supplementary Material.

Doping increasingly separates the
QP-dip-hump features at the Fermi energy
from the region of band inversion between the bottom of the CB and the top of the MVB,
confirming that they are distinct, mutually independent characteristics. In particular SVBs are clearly distinct from topological  surface states.
The doping dependence of the ARPES spectra also
rules out an interpretation of the SVBs in terms of states bound
to a surface layer detached from the bulk
(see Supplementary Material for further discussion).

Finally, doping reveals that,
while CB and MVB are shifted by roughly the same energy at
$\hbox{t1}\equiv\hbox{1\,minutes}$,
their shift differs from the nearly rigid energy shift of the lower VBs.
This suggests that the lower VBs 
are nearly uncorrelated with the carriers
at the Fermi level, while  MVB and  CB are  more strongly correlated
with new doped carriers. This may be the reason why we observe excitons
only associated with the MVB, but not with the other VBs,
(see Sec.\ III%\ref{appsubsec: Doping dependence of ARPES spectra}
of the Supplementary Material).

\textbf{Interpretation of SVBs}
As is explained in the Supplementary Material and summarized
in Fig.\ \ref{Fig: 4},
the position, shape, and doping dependence of the SVBs
can best be explained in terms of a moving bound state between
the photo-hole with non-zero group velocity and large momentum
in the VB of a given chain
and one (or possibly several) QPs in the CB on the same or on
neighboring chains, see Figs.\ \ref{Fig: 4}\textbf{a-b}. In particular,
for sufficiently large total momentum $K$ along the chain (as compared to
$\Delta k$, the diameter of the polaronic islands),
we propose the following interpretation. 
The MVB arises from the excitation of a single photo-hole
in the VB, see Fig.\ \ref{Fig: 4}\textbf{cI}.
The top-most SVB above the MVB is the $K$-dependent threshold to a continuum
consisting of an exciton and a free quasihole in the CB,
the exciton being a moving bound state
between a QP in the CB and a valence hole
on the same chain, sharing the same group velocity
Fig.\ \ref{Fig: 4}\textbf{cII}.
The SVBs closer to the MVB can be of two
distinct origins. One possibility is that the particle-hole excitation
in the CB is created on a chain neighboring the one hosting
the valence hole, leading to an interchain exciton
Fig.\ \ref{Fig: 4}\textbf{cIII},
with lower binding energy than intrachain excitons. 
Alternatively, one can have thresholds to more complex continua,
the simplest one consisting in a photo-hole accompanied by two particle-hole
excitations on neighboring chains
Fig.\ \ref{Fig: 4}\textbf{cIV},
whereby both quasiparticles bind
to the valence hole to form a mobile trion
\cite{Lampert58,Kheng93,Matsunaga11}.
These trions are distinct from two-particle bound states (excitons)
that are dressed with particle-hole excitations of the Fermi sea, occuring in optical low-momentum excitations
of monolayer transition-metal dichalcogenides
at finite carrier densities in Ref.\ \onlinecite{Efimkin17}.
A trion threshold is expected
to have a lower energy-integrated intensity, as it scales with a higher power
of the density of conduction electrons
%DROP?
(since it involves two particle-hole excitations),
an aspect that might rationalize the low intensity
of the second SVB in Fig.\ \ref{Fig: 3}\textbf{c-d}.

The separation of the SVBs from the MVB results from two sources,
the binding energy and the kinetic energy gain from distributing the
momentum between the VB and the CB.
We show with Fig.\ \ref{Fig: 4}\textbf{b}
how these two effects combine to produce SVBs above the MVB.
The binding energy is relatively strong due to the heavy mass of the
polaronic quasiparticles and the effective one-dimensionality (which
entails an increase of binding energy with doping),
as we discuss in
Secs.\ VI-VIII
%\ref{appsubsec: Mobile excitons in a one-dimensional metal}-%
%\ref{appsubsec: Doping dependence of 1D exciton binding energies}
in the Supplementary Material.
In a pristine sample we estimate it to be of the order of
$150\,\mathrm{meV}$ for the most strongly bound exciton
% (Sec.\ V \ref{appsubsec: Mobile excitons in a one-dimensional metal}
%in the Supplementary Material).

While the polaronic reconstruction of quasiparticles makes the
formation of excitons more robust, excitonic excitations can also occur
without polaronic reconstruction, provided the
%the latter can also occur if the
Fermi surface is sufficiently small.

\textbf{Discussion}

The material TaSe${}_{3}$ is a Q1D metal which
combines a low density of conduction electrons with a polaronic
renormalization of the low-energy quasiparticles. All these ingredients
{enhance} the spectral weight in photoemission for
composite excitations involving an exciton and a hole from the CB.
Our experiments show that the excitons come with different internal structure,
presumably depending on whether the involved holes and electrons belong to the
same chain or neighboring ones, or whether the hole binds one or two
conduction electrons (resulting in an exciton or a trion, respectively).

Interchain excitons are quasi-1D cousins of
bilayer excitons in layered 2D materials,
such as transition metal dichalcogenides\,
\cite{Rivera2015,Wilson2017}.
They are of particular interest as they may have a significantly longer
life time than intra-chain excitons due to the spatial separation of
the particle and the hole.

We have experimentally probed the evolution of these SVBs with
increasing doping. The latter increases both the binding energy
(owing to the Q1D nature of the problem) and the typical momentum transfer
in the CB. Both increase the energy separation
between the SVBs and the MVB.  A more systematic
study of the doping dependence will allow to analyze the fate and
nature of excitons as one crosses the Lifshitz transition where a
second CB emerges at the Fermi level. It would also be
interesting to excite the exciton bound states at low momenta by
optical absorption, or by using resonant inelastic X-ray scattering,
or exciting them by means different from light. Finally,
it would be interesting to see whether dispersing excitons also exist in other
trichalcogenides, at least in the regime of higher temperatures where
they do not form a charge density wave. 
This will elucidate to what extent the peculiar
structure of the polaronic band with its strong intensity islands is
crucial for the visibility of exciton branches in ARPES.

%%% so far 3700 words.

\textbf{Acknowledgments}
We acknowledge E. Rienks, H. Li, Y. Hu, and V. Strokov for help during
the ARPES experiments. We thank Prof.\ Dirk van der Marel for discussions.
M.S., J.Z.M. and J.J. were supported by the Sino-Swiss Science and
Technology Cooperation (Grant No.\ IZLCZ2-170075). M.S. was supported by the Swiss National Science Foundation under Grant No. 200021$\_$188413. M.S., O.Y. and D.G. were supported by the NCCR MARVEL funded by the Swiss National Science Foundation.
M.M. was supported by the Swiss National Science Foundation under
Grant No.\ 200021$\_$166271. 
M.N. has received funding from
the European Union's Horizon 2020 research and innovation programme
under the Marie Sklodowska-Curie grant agreement No.\ 701647,
and Swiss National Science Foundation under Grant 200021$\_$159678.
J.Z.M. was supported by City University of Hong Kong through
the start-up project (Project No.\ 9610489), the National Natural Science Foundation of China (12104379), 
and Shenzhen Research Institute, City University of Hong Kong.
M.A.S. is supported by Deutsche Forschungsgemeinschaft
through the Emmy-Noether programme
is gratefully acknowledged (SE 2558/2-1). W.W.X. was supported by
Beckman Young Investigator Program funded by Arnold and Mabel Beckman
Foundation and US NSF DMR-1944965. X.G. was supported by the US
Department of Energy Division of Basic Energy Sciences
(DG-FG02-98ER45706).

\textbf{Author contributions:}
J.Z.M. performed ARPES experiments with the help
of M.N., J.J. and W.H.F.; S.M.N. and Z.W. performed first-principles
calculations of the band structure. J.Z.M. plotted all the figures.
X.G. and W.W.X. synthesised the
single crystals. C.Y.X. performed primary high magnetic field quantum
oscillation (QO) measurements with the help of J.Z.M., J.L.Z.,
T.S. and Y.M.X.; H.H. and U.D.G. analyzed the possibility of boson
driven band structures with the help of D.G. and O.Y.; I.K. performed
Raman measurements for checking the phonon energy. N.K. and
Y.S. performed primary transport measurements with
PPMS; M.A.S.  helped ruling out the bosonic strong coupling scenario.
C.M. and M.M.  analyzed different physical scenarios based on bound
states and worked out the theory of excitonic side-bands. All authors
contributed to the discussion of this project.  J.Z.M., C.M., M.M., and
M.S. wrote the manuscript.

\textbf{Competing interests:} The authors declare that
they have no competing interests.

\begin{figure*}[t]
\begin{center}
\includegraphics[angle=0,width=0.9\textwidth]{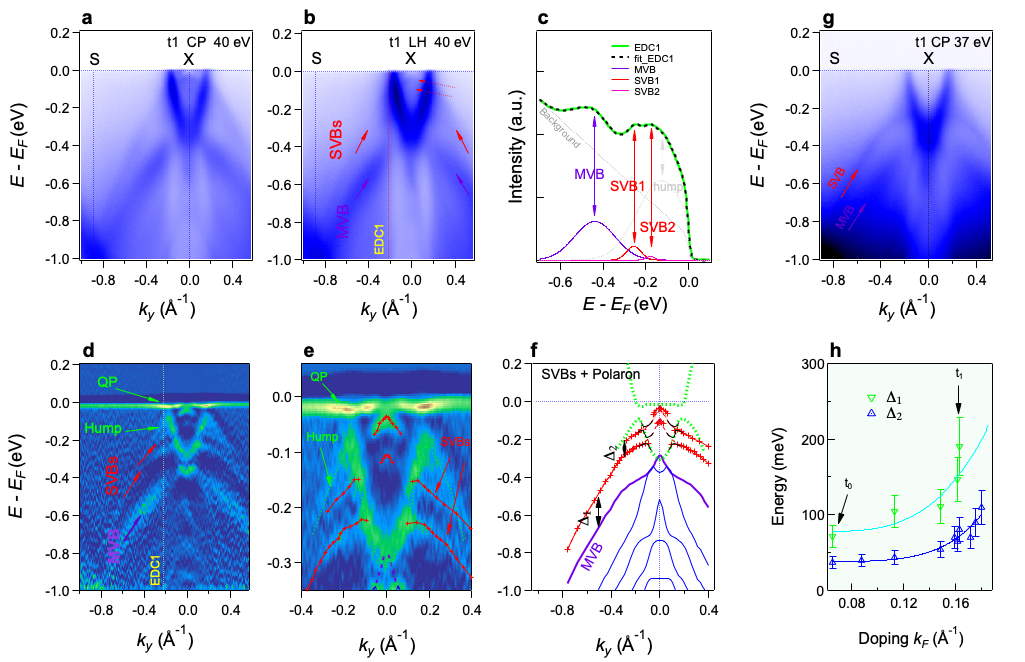}
\end{center}
\caption{%
%\scriptsize
Dependence on potassium doping of the electronic structure of TaSe${}_{3}$.
\textbf{a-b} ARPES intensity plot with energy measured relative to
$E^{\,}_{\mathrm{F}}$ and momenta along the X-S cut,
after potassium was deposited for t1=1 minute on the surface of TaSe${}_{3}$.
The incoming photons are circular positive-helicity [CP,\textbf{a}]
and linearly horizontal [LH, \textbf{b}] polarized,
respectively. The full red arrows in \textbf{b} indicate a
side valence band (SVB) above the main valence band (MVB),
to which the purple arrow points.
The dashed red arrows indicate additional spectral side-features at small
$k^{\,}_{y}$, also seen in \textbf{a}.
\textbf{c}
Fit of the EDC1 in panel \textbf{b}:
MVB, SVB1, SVB2, and hump are modeled as Gaussians,
superposed on a linear background, and cut off by the Fermi-Dirac function.
Two side peaks, a strong one (SVB1) and weak one (SVB2),
are visible at that value of $k^{\,}_{y}=-0.2\,\text{\AA}^{-1}$. 
\textbf{d} The curvature intensity plot of the ARPES intensity in
\textbf{b}.
The two SVBs are indicated by red arrows.
The MVB peak is indicated by a purple arrow.
The coherent polaron peak (QP) and the bare CB excitation (hump)
are indicated by green arrows. 
\textbf{e} The close up of the curvature intensity plot in \textbf{d}
identifies the SVBs more clearly. The red lines show
the two SVB dispersions extracted from the peak positions
in the curvature intensity plot.
Additionally we delineate nose-like side spectral features at small $k^{\,}_{y}$.
\textbf{f} The band dispersions extracted from tracing the peak positions of
the full spectrum in \textbf{d}. The green dotted lines are the polaronic CB
and the dispersing hump of the bare CB excitation.
The red lines at small $k^{\,}_{y}\approx 0$ 
are sidepeaks whose vertical shift relative to the SVBs at large momenta
is similar as that of the ``nose''-like peak predicted by the DFT calculations
of Fig.\ \ref{Fig: 2}\textbf{a}
with respect to the MVB 
dispersion at large $k^{\,}_{y}$.
The dotted red lines suggest that these side-features
might be of similar origin and form a band,
which is however interrupted by avoided crossings
(indicated by thin black lines)
with the bare CB (hump).
\textbf{g} ARPES intensity as a function of energy relative to
$E^{\,}_{\mathrm{F}}$ and momentum $k^{\,}_{y}$
along the X-S cut recorded with
$h\,\nu=37\,\mathrm{eV}$
after potassium was deposited for 1 minute on the surface of TaSe${}_{3}$.
\textbf{h}
The average energy separations $\Delta^{\,}_{1}$
(between MVB and the first SVB) and $\Delta^{\,}_{2}$
(between the first and the second SVB), (see panel \textbf{f}),
plotted as a function of the measured Fermi wave vector $k^{\,}_{\mathrm{F}}$,
which is tuned by the evaporation time of potassium.
All the data in fig.3 is recorded at 19K.
       }
\label{Fig: 3}
\end{figure*}

\begin{figure*}[t]
\begin{center}
\includegraphics[angle=0,width=0.6\textwidth]{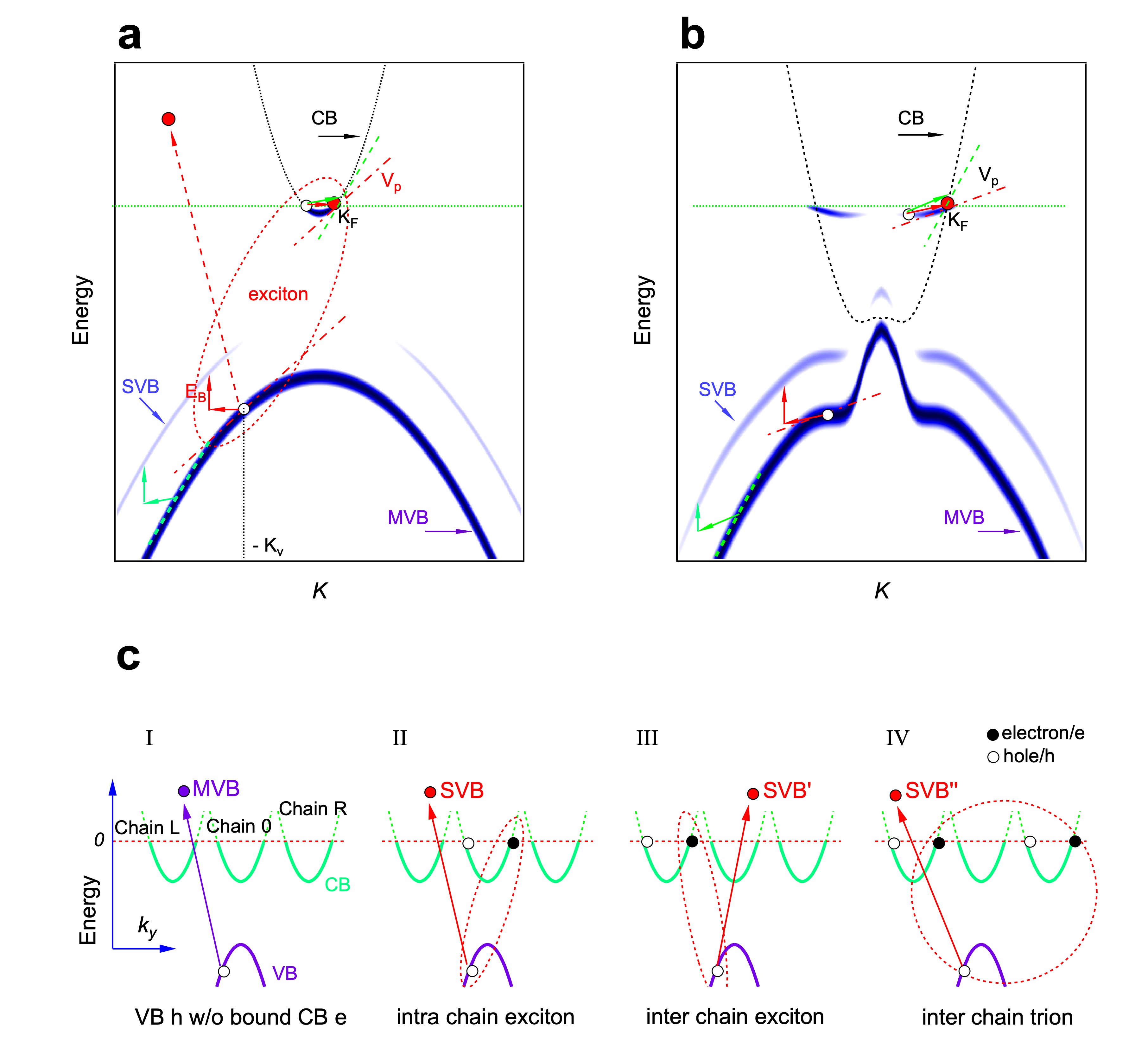}
\end{center}
\caption{%
Schematic of how mobile excitons show up in the form of SVBs in photoemission.
\textbf{a} Toy model with parabolic VB and CB.
The drawing shows excitonic excitations for total
momenta $K>K^{*}$ where a VB hole with velocity
$v^{\,}_{\mathrm{h}}>v^{\,}_{\mathrm{F}}$
binds to a CB electron above the Fermi level. The SVB is constructed
graphically{.} Starting from a VB state, the partner state with equal
group velocity (slope of the dispersion) in the CB is determined. The
energy-momentum transfer from the left Fermi point to that state is
combined with a vertical shift by the (momentum independent)
binding energy to obtain a point on the excitonic side branch.
\textbf{b} The same procedure applied to a more realistic model of
TaSe${}_{3}$. The simple parabolic Fermi sea is replaced by high
intensity islands in the polaronic band, its diameter $\Delta k$
playing the role of $2k^{\,}_{\mathrm{F}}$ in \textbf{a}.
The binding energy now depends on
the curvature of the VB and the CB at the momenta involved in the
exciton pair. The SVB construction results in an excitonic
branch which is a nearly parallel translate of the simple hole
dispersion, the MVB.
\textbf{c} Proposed origin of the MVB and the various SVBs{.
T}he MVB \textbf{I} results from a single VB hole excitation.
It costs less energy (at given momentum $K$) if it is
accompanied by particle-hole excitations in the CB of the same or
neighboring chains. The binding of the CB particle(s) with the VB hole
results in various types of excitonic modes,
namely intrachain exciton \textbf{II},
interchain excitons \textbf{III}, and possibly interchain trions
\textbf{IV}.
         }
\label{Fig: 4}
\end{figure*}

%\bibliographystyle{acm}
%\bibliography{references}

\textbf{Methods}

\textbf{Crystal synthesis and structure}.
Single crystals of TaSe${}_{3}$ were synthesized
using evenly-ground stoichiometric elemental Ta
($\sim325$ mesh, Beamtown Chemical, $\geq99.9\%$) and Se
($\sim200$ mesh, Beamtown Chemical, $\geq99.999\%$)
with a total mass of $\sim\mathrm{300mg}$ were pressed
into a pellet and placed in an evacuated quartz tube mixed with
$\sim\mathrm{10mg}$
of I${}_{2}$ as the vapor transport medium. The tube was put into a tube
furnace and heated up to $\mathrm{700}\degree\mathrm{C}$
on the sample side while the other
side of the tube,
while in the furnace, was open to the external environment so as
to generate a temperature gradient. After holding the sample at
$\mathrm{700}\degree\mathrm{C}$ for five days,
the furnace was shut down and cooled to room
temperature. Needle-like crystals TaSe${}_{3}$ can be found at the cooler side of
the quartz tube.
The trichalcogenide TaSe${}_{3}$ is a representative of the
family of crystals XT${}_{3}$ where X belongs to either the
group IVB (Ti, Zr, Hf) or the group VB (Nb, Ta),
and T is a chalcogen from the group VIA (S, Se, Te).
The crystalline structure of
TaSe${}_{3}$ is monoclinic with
the space group P2${}_{1}$/m (SG11).
The selenium atoms are located at the vertices of
triangular prisms with three faces parallel to the $b$-axis and
a tantalum atom  at their center.
These prisms are covalently stacked along the
$b$-axis and form parallel one-dimensional (1D)
chains.
The unit cell viewed along the $b$-axis
contains four triangular prisms,
with covalent bonds along the $(101)$ direction.
Thus, the natural cleavage plane is the $(10\bar{1})$
surface.

\textbf{ARPES experiments}.
 ARPES measurements were performed at the SIS-ULTRA beamline
  and at the ADRESS-ARPES beamline of the Swiss Light Source, Paul Scherrer Institute,
and at the beamline UE112 PGM-2b-1${}^{3}$ at BESSY (Berlin Electron Storage
Ring Society for Synchrotron Radiation) Synchrotron. The energy and
angular resolutions were set to 
$5-30\,\mathrm{meV}$ and \ang{0.1},
respectively. The
samples for ARPES measurements were cleaved in situ and measured mainly in a
temperature range between $\mathrm{15K}$ and $\mathrm{25K}$
in a vacuum exceeding $8\times10^{-11}\mathrm{Torr}$.

\textbf{First-Principles calculations}.
The first-principles calculations were performed within
the framework of the full-potential linearized augmented plane-wave
method implemented in the WIEN2K simulation package. A modified
Becke-Johnson exchange potential together with the local-density
approximation for the correlation potential was used to obtain
accurate band structures. Spin-orbit coupling (SOC) was included as a second,
self-consistent variational step. The radii of the muffin-tin sphere
$R^{\,}_{\mathrm{MT}}$ were 2.5 Bohr
for Ta and 2.38 Bohr for Se. The $\bm{k}$-point sampling
grid of the BBZ in the self-consistent process was $7\times19\times6$. The
truncation of the modulus of the reciprocal lattice vector
$K^{\,}_{\mathrm{max}}$, which
was used for the expansion of the wave functions in the interstitial
region, was set to
$R^{\,}_{\mathrm{MT}}\,K^{\,}_{\mathrm{max}}$=7.
The geometry optimization including
SOC was carried out within the framework of the projector
augmented-wave pseudopotential method implemented in the Vienna Ab initio
Simulation Package. The ionic positions were relaxed
until the force on each ion was less than $0.005\,\mathrm{eV}/\text{\AA}$.

\textbf{Data Availability}.
The data that supports the findings of this study is available in the MARVEL public repository (MARVEL Materials Cloud Archive: https://archive.materialscloud.org) with same title of this paper.

\end{document}